\documentclass[12pt,floats]{article}
\usepackage{epsfig}
\begin{document}

\title{\bf Proposition of numerical modelling of BEC}

\author{Oleg V. Utyuzh$^{1}$, Grzegorz Wilk$^{2}$ and
Zbigniew W\l odarczyk$^{3}$
\\[2ex]
$^{1}${\it Nuclear Theory Department,} \\
      {\it The Andrzej So\l tan Institute for Nuclear Studies,}\\
      {\it Ho\.za 69, 00681 Warsaw, Poland,} \\
      {\it Tel: +48 22/ 553 22 28, Fax: +48 22/ 621 60 85,} \\
      {\it e-mail: utyuzh@fuw.edu.pl }\\[1ex]
$^{2}${\it Nuclear Theory Department,} \\
      {\it The Andrzej So\l tan Institute for Nuclear Studies,}\\
      {\it Ho\.za 69, 00681 Warsaw, Poland,} \\
      {\it Tel: +48 22/ 553 22 26, Fax: +48 22/ 621 60 85,} \\
      {\it e-mail: wilk@fuw.edu.pl }\\[1ex]
$^{3}${\it Institute of Physics,} \\
      {\it \'Swi\c{e}tokrzyska Academy,} \\
      {\it \'Swi\c{e}tokrzyska 15, 25-405 Kielce, Poland,} \\
      {\it Tel./Fax: +48 41/ 362 64 52,} \\
      {\it e-mail: wlod@pu.kielce.pl }}

\date{\today}
\maketitle

\begin{abstract}
We propose extension of the numerical method to model effect of
Bose-Einstein correlations (BEC) observed in hadronization processes
which allows for calculations not only correlation functions
$C_2(Q_{inv})$ (one-dimensional) but also corresponding to them
$C_2(Q_{x,y,z})$ (i.e., three-dimensional). The method is based on the
bunching of identical bosonic particles in elementary emitting cells
(EEC) in phase space in manner leading to proper Bose-Einstein form of
distribution of energy (this was enough to calculate $C_2(Q_{inv})$). To
obtain also $C_2(Q_{x,y,z})$ one has to add to it also symmetrization of
the multiparticle wave function to properly correlate space-time
locations
of produced particles with their energy-momentum characteristics.\\

\end{abstract}

Recently we have proposed novel numerical method to model effect of the
Bose-Einstein correlations (BEC) observed in hadronization processes (see
\cite{HIP,NUKL} for details and for other references concerning
generalities of BEC). This method allows for calculation of the so called
invariant (i.e., one-dimensional) correlation function $C_2(Q_{inv})$.
Referring to \cite{HIP,NUKL} for motivation and justification and all
other details let us list here only the main points proposed so far:
\begin{itemize}

\item Bosonic character of produced secondaries demands that they are
produced in bunches (named by us {\it Elementary Emitting Cells} - EEC's)
in momentum space \cite{Purcell}.

\item The problem then is how to model production of EEC. It is done by
adding to some preselected particle with energy $E_1$ (according to some
distribution $f(E)$, which in our numerical calculations was taken in the
Bolztmann form, $f(E) \sim \exp(-E/T)$, with temperature $T$ being a
parameter) another particles of the same energy with some probability
$P=P_0\cdot \exp(-E_1/T)$ (where $P_0$ is another parameter) up to first
failure. After it one starts to build another EEC and continues as long
as total energy allows.

\item Such procedure results in a number of EEC's (distributed according
to Poisson distribution and follow boltzmanian energy level distribution)
with a number of identical particles in each of them distributed
according to geometrical (Bose-Einstein) distribution with energy level
distribution following Bose-Einstein statistics.

\item In this way one gets $C_2(Q_{inv})=2$ but only in the first bin,
i.e., for $Q_{inv}=0$. In order to get the characteristic shape of
$C_2(Q_{inv})$ function one has to allow for particles in each EEC to
have slightly different energies, for example distributed around $E_1$
according with gaussian distribution with the width $\sigma$, which is
our next and last parameter on this stage. It should be stressed at this
point that in the field theory approach to BEC, as for example that
presented in \cite{Kozlov}, $\sigma=0$ corresponds to infinite source and
commutation relations with Dirac-delta functions, whereas nonzero values
of sigma arise for finite space-time extensions of the hadronizing
sources.

\end{itemize}

To obtain also correlation functions $C_2(Q_{x,y,z})$ ( i.e.,
three-dimensional, calculated for different components of the differences
of particle momenta $Q$)one has to proceed further. What we propose here
is the following:
\begin{itemize}

\item Change of $Q_{inv}$ to $\vec{Q}$ means that we shall be now
sensitive not only to overall spatial difference between particle
production points $r$ but to the whole vector $\vec{r}$ as well. We have
to then assume that particles are produced from some spatial region and
that density of production points is given by some function (our
additional input) $\rho(x,y,z)$ \footnote{For a time being we are
assuming for simplicity that hadronization is instantaneous so $\rho$ is
time-independent.}

\item Momenta of each particle in a given EEC, $p_i$, obtained in the
first part of algorithm must now be decomposed into their components,
$\left(p^{(i)}_x, p^{(i)}_y, p^{(i)}_z\right)$. To do this let us observe
that BE statistics demands that multiparticle wave functions must be
symmetrized accordingly and this results in correlations between
production points represented by $\rho(x,y,z)$ and momenta
$\left(p^{(i)}_x, p^{(i)}_y, p^{(i)}_z\right)$. Denoting by
$\delta_{i=x,y,z}$ the corresponding differences in position this
correlation is given (in the plane wave approximation) by the known
$1+\cos\left(\delta_i\cdot Q_i \right)$ term.

\item We proceed then in the following way. In each EEC we select the
$\left(p^{(i)}_x, p^{(i)}_y, p^{(i)}_z\right)$ for the first, $i=1$
particle in some prescribe way (here isotropically but one can introduce
at this moment $p_T$ cutting or something else as well) and then
establish $\left(p^{(i)}_x, p^{(i)}_y, p^{(i)}_z\right)$ for every one of
additional particles, $i\ge 2$, in such a way that $\cos(\delta_i \cdot
Q_i) \le 2\cdot Rand -1$ where $Rand$ is random number uniformly chosen
from interval $(0,1)$.

\end{itemize}
This leads us to results presented in Figs. 1 and 2. We regard them as
very promising but we are aware of the fact that our proposition is still
far from being complete. To start with one should allow for time
depending emission by including $\delta E\cdot \delta t$ term in the
$\cos(\dots)$ above. The other is the problem of Coulomb and other final
state interactions. Their inclusion is possible by using some distorted
wave function instead of the plane waves used here. Finally, so far only
two particle symmetrization effects have been accounted for: in a given
EEC all particles are symmetrized with the particle number $1$ being its
seed, they are not symmetrized between themselves. To account for this
one would have to add other terms in addition to the $\cos(\dots)$ used
above - this, however, would result in dramatic increase of the
calculational time.


\begin{center}
\begin{tabular}{|cc|}
\hline
\begin{minipage}{7cm}
\begin{center}
\includegraphics[height=4.5cm,width=7cm]{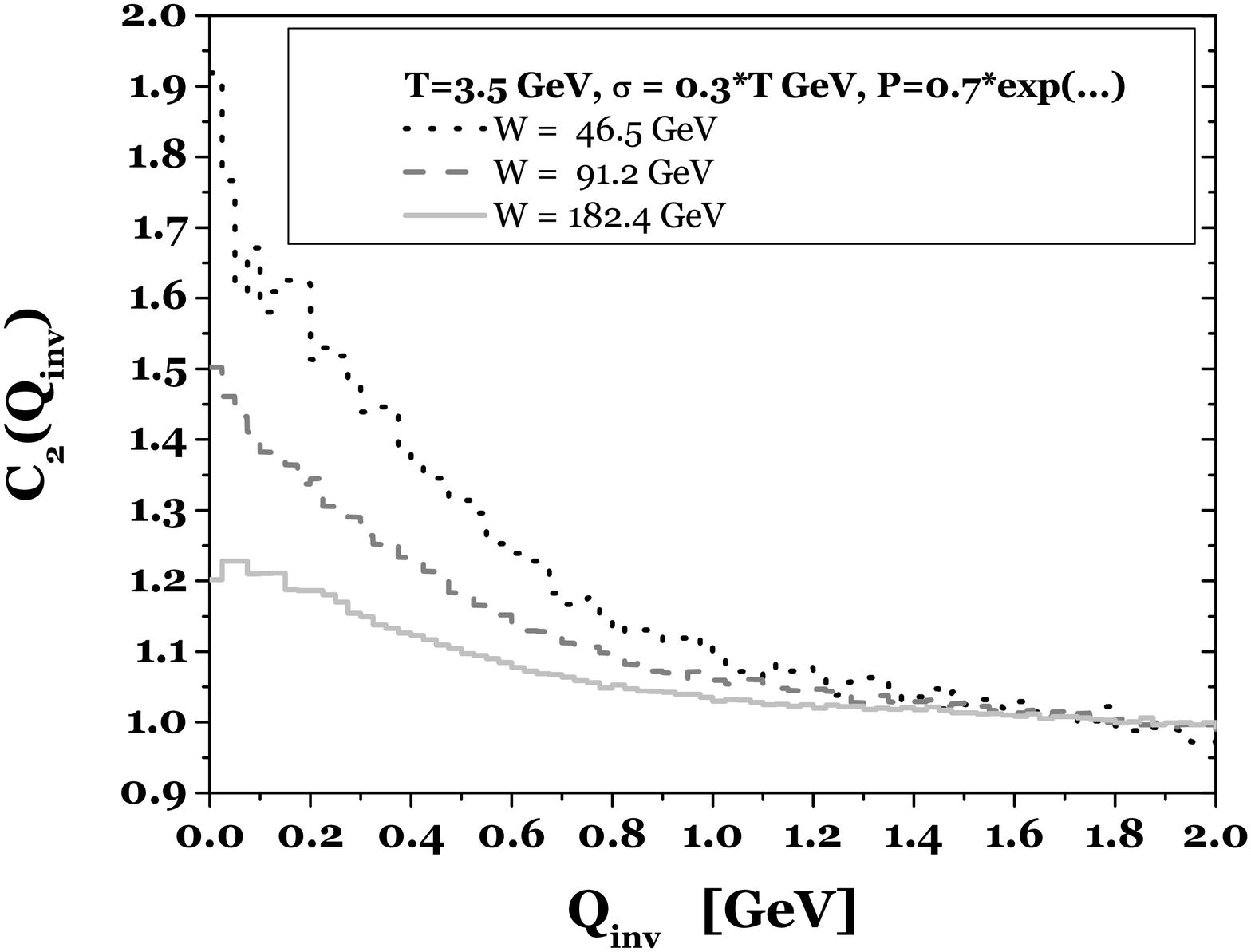}
\end{center}
\end{minipage} &
\begin{minipage}{7cm}
\begin{center}
\includegraphics[height=4.5cm,width=7cm]{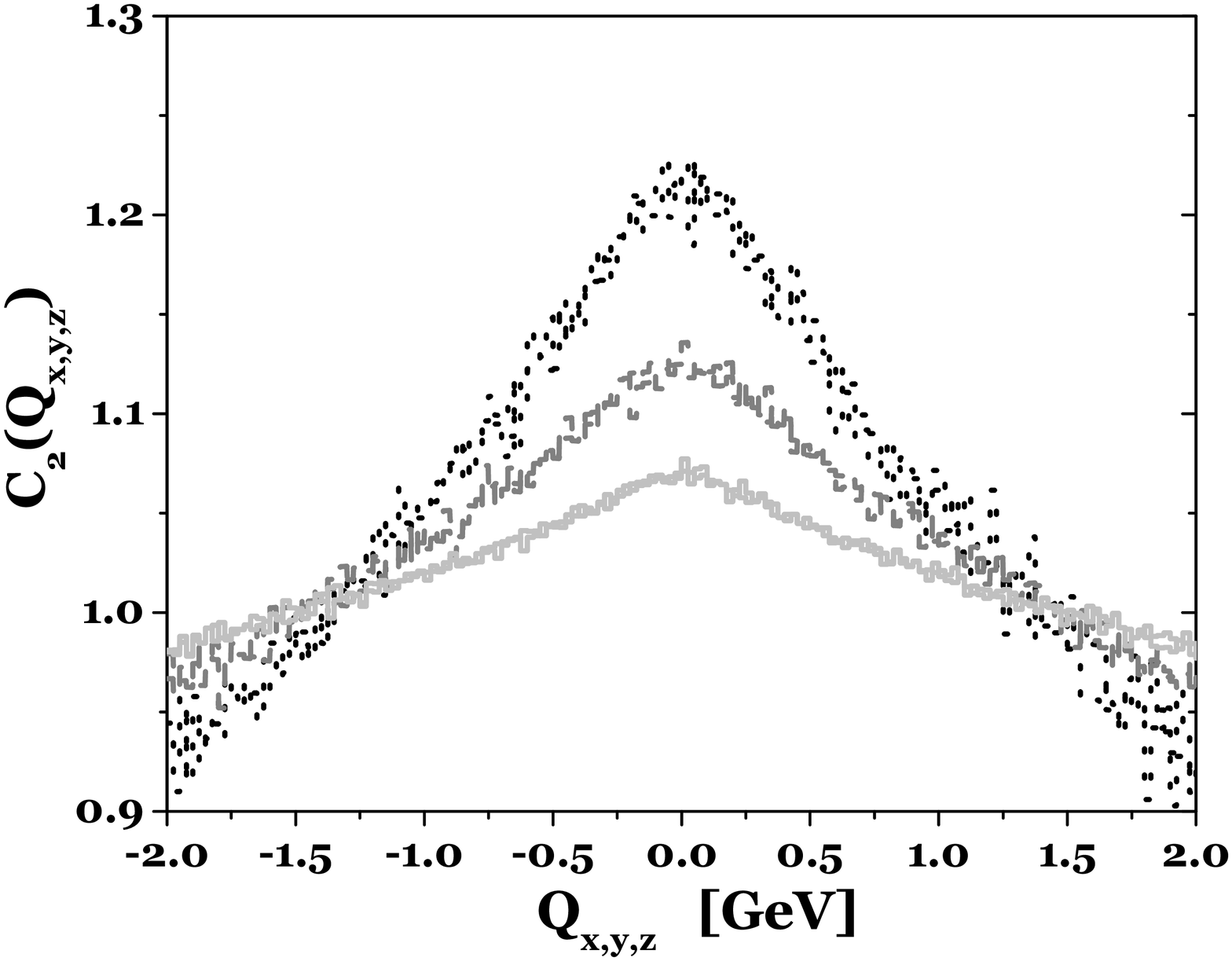}
\end{center}
\end{minipage} \\
\begin{minipage}{7cm}
\begin{center}
\includegraphics[height=4.5cm,width=7cm]{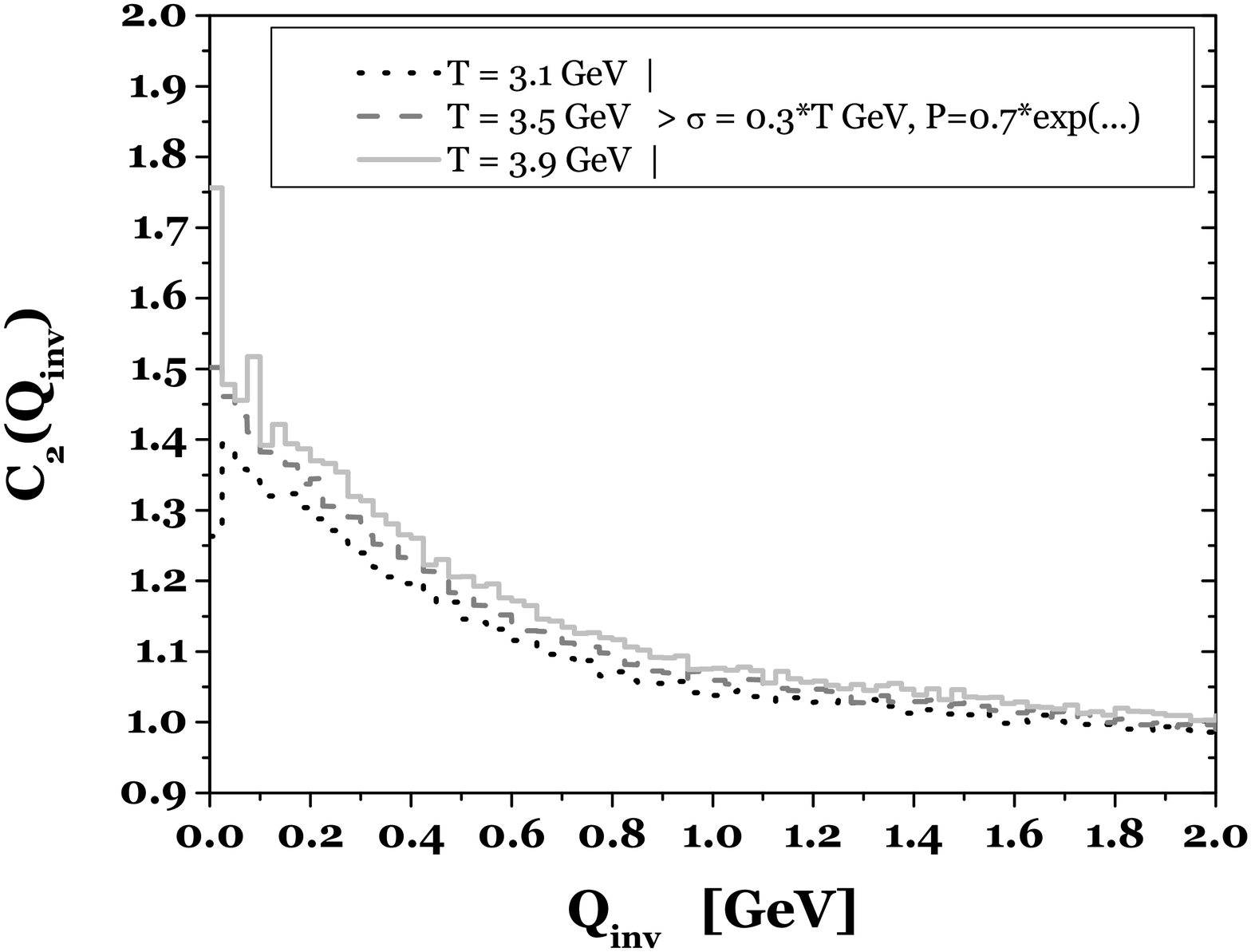}
\end{center}
\end{minipage} &
\begin{minipage}{7cm}
\begin{center}
\includegraphics[height=4.5cm,width=7cm]{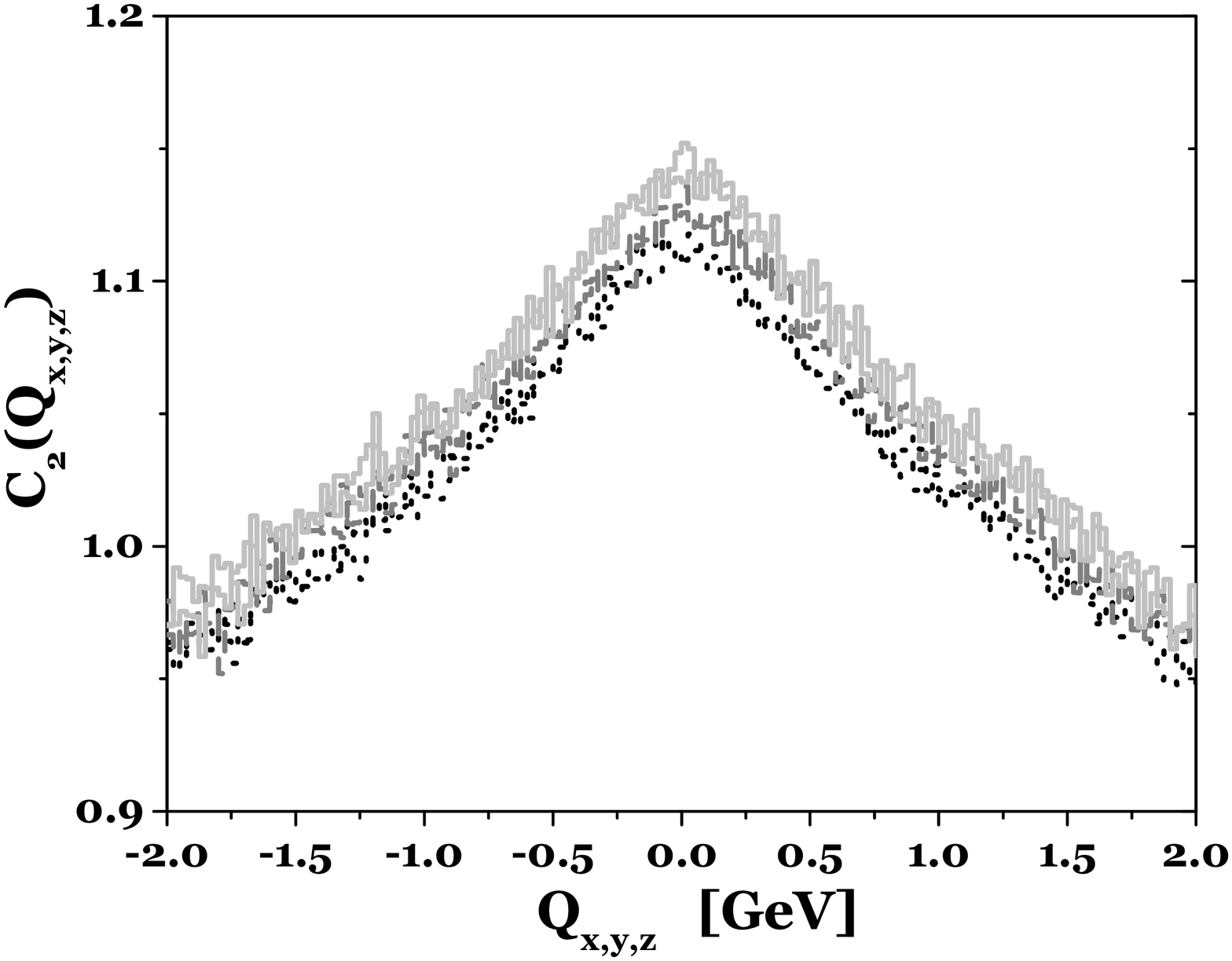}
\end{center}
\end{minipage} \\
\begin{minipage}{7cm}
\begin{center}
\includegraphics[height=4.5cm,width=7cm]{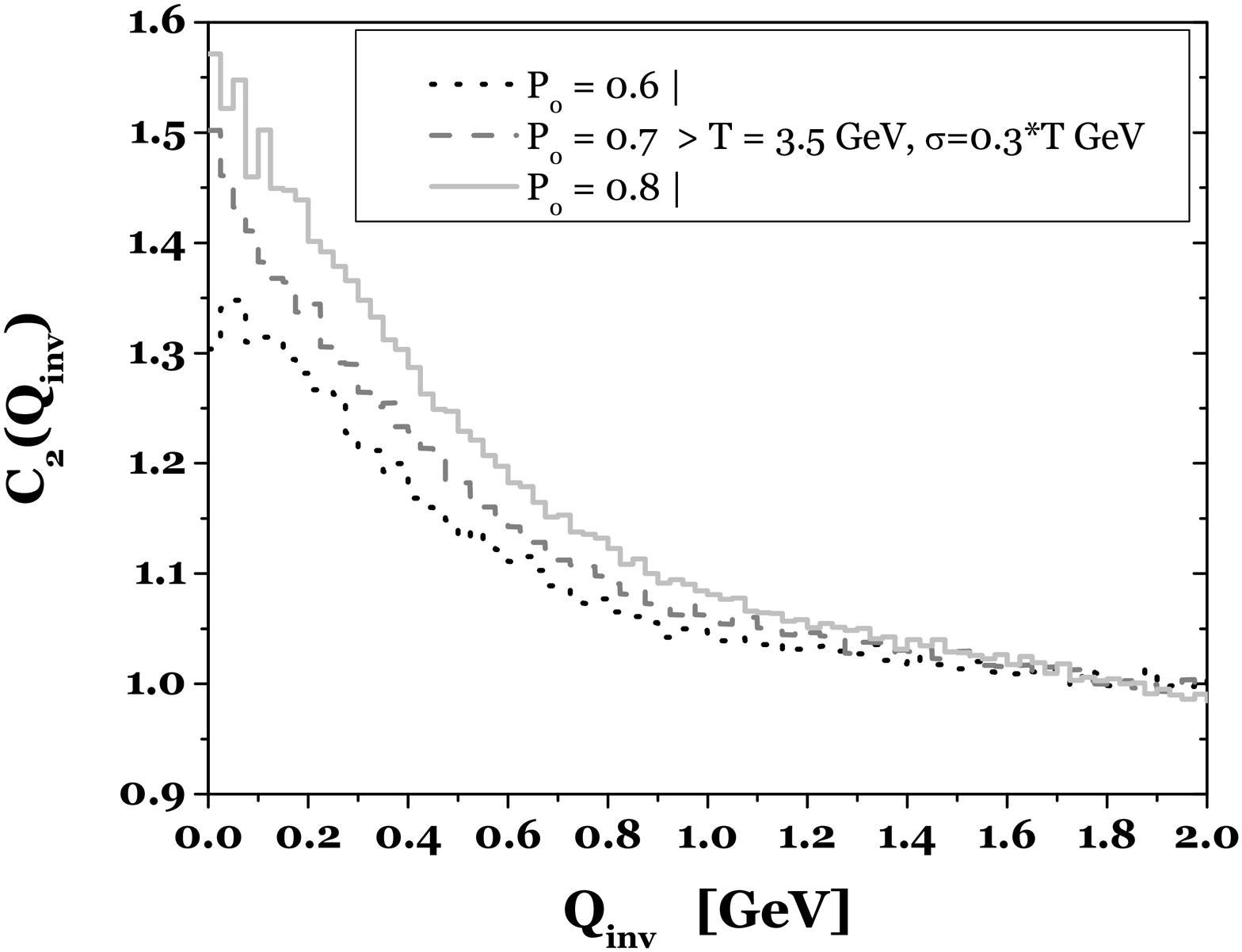}
\end{center}
\end{minipage} &
\begin{minipage}{7cm}
\begin{center}
\includegraphics[height=4.5cm,width=7cm]{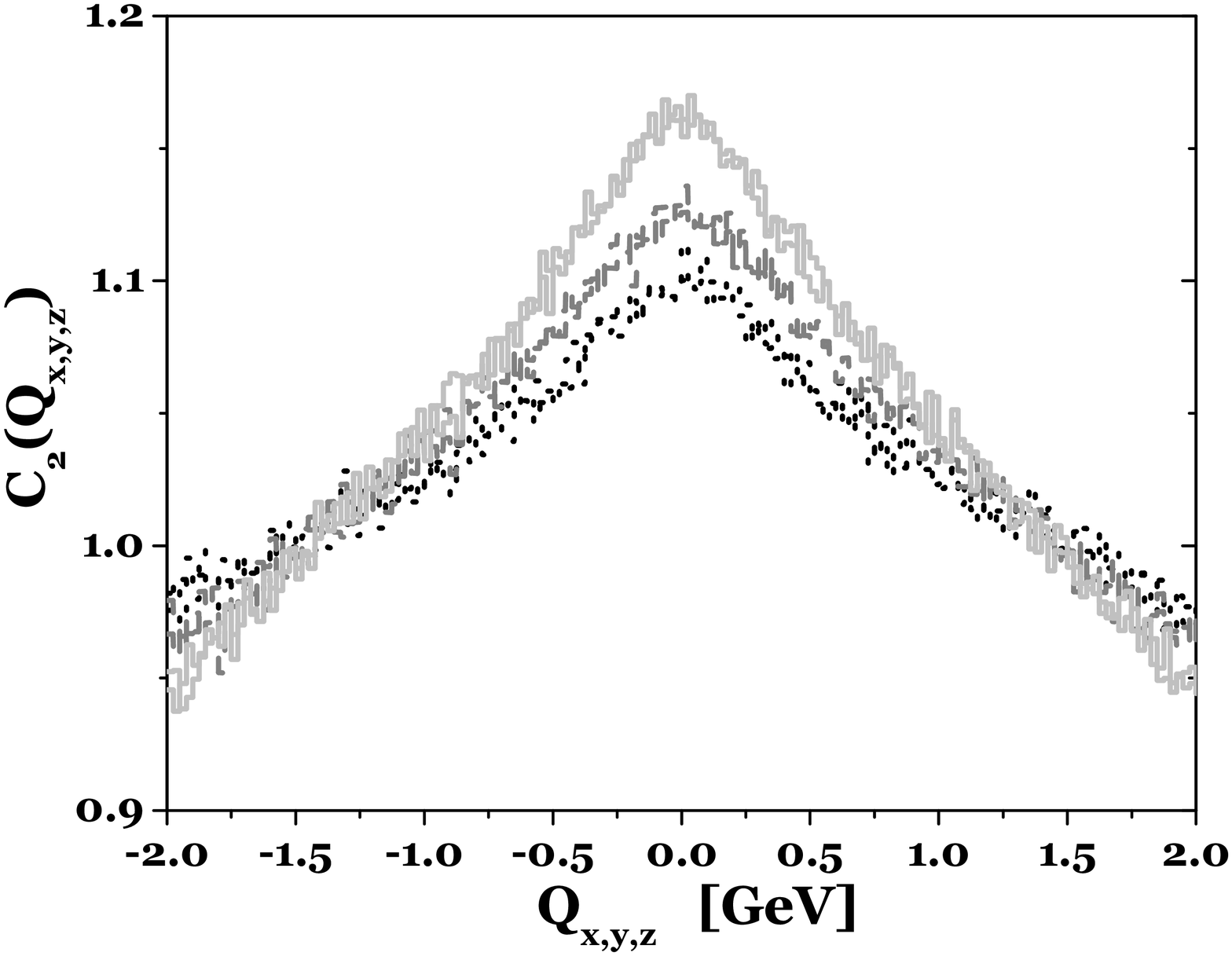}
\end{center}
\end{minipage}  \\
\begin{minipage}{7cm}
\begin{center}
\includegraphics[height=4.5cm,width=7cm]{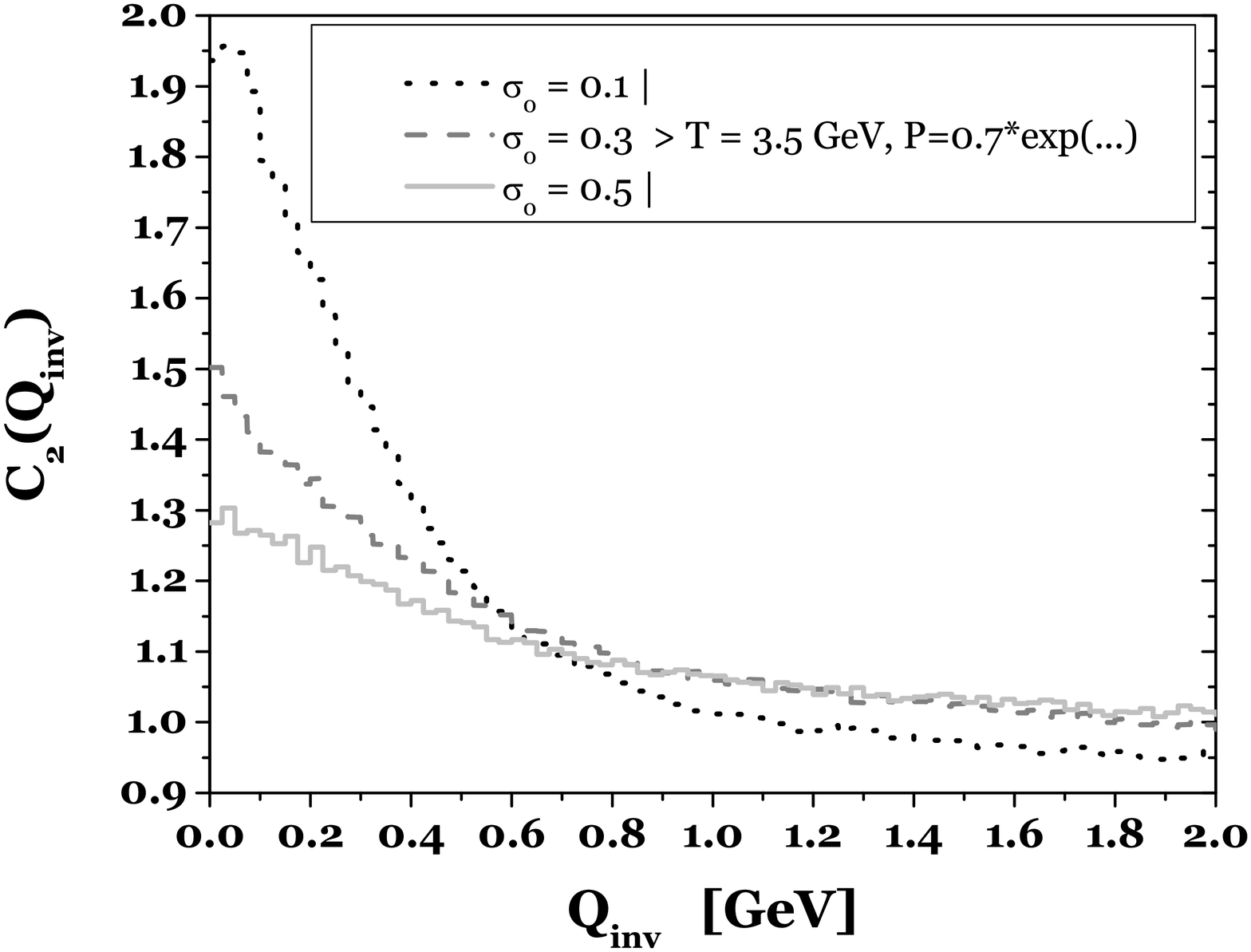}
\end{center}
\end{minipage} &
\begin{minipage}{7cm}
\begin{center}
\includegraphics[height=4.5cm,width=7cm]{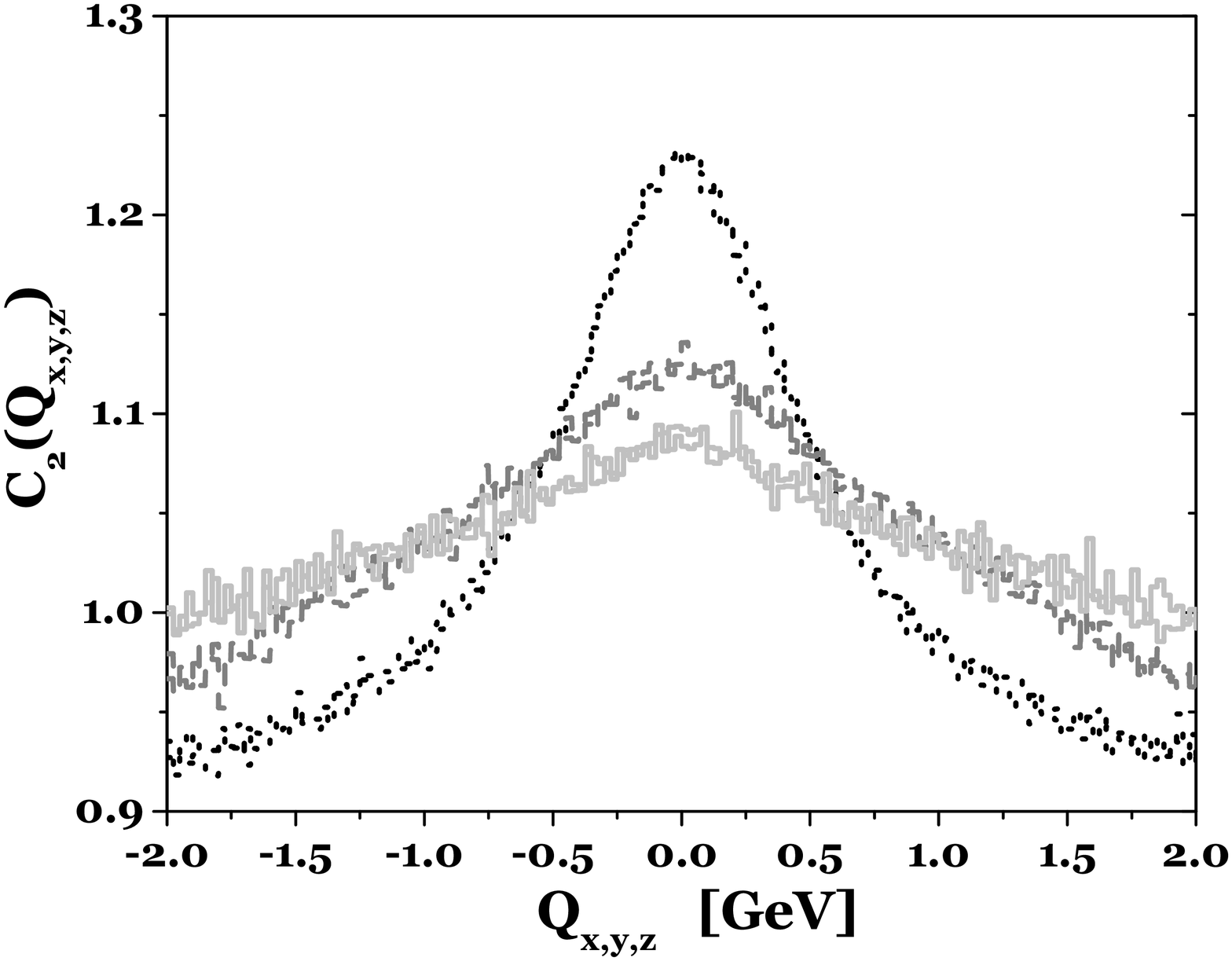}
\end{center}
\end{minipage}  \\
\hline
\end{tabular} \\
\vspace{0.5cm}{\scriptsize Fig.1 Example of results obtained for
$C_2(Q_{inv})$ (left column) and corresponding $C_2(Q_{x,y,z})$
(right column, for parameters used here all $C_2(Q_{x,y,z})$ are the
same). Calculations were performed assuming spherical source
$\rho(r)$ of radius $R=1$ fm (spontaneous decay was assumed,
therefore there is no time dependence) and spherically symmetric
distribution of $p_{x,y,z}$ components of momenta of secondaries
$p$. Energies were selected from $f(E) \sim \exp( -E/T)$
distribution. The changes investigated are - from top to bottom:
different energies of hadronizing sources, different temperatures
$T$, different values of parameter $P_0$ and different spreads
$\sigma=\sigma_0\cdot T$ of the energy in EEC.}.
\end{center}

 \begin{center}
\begin{tabular}{|cc|}
\hline
\begin{minipage}{7cm}
\begin{center}
\includegraphics[height=4.5cm,width=7cm]{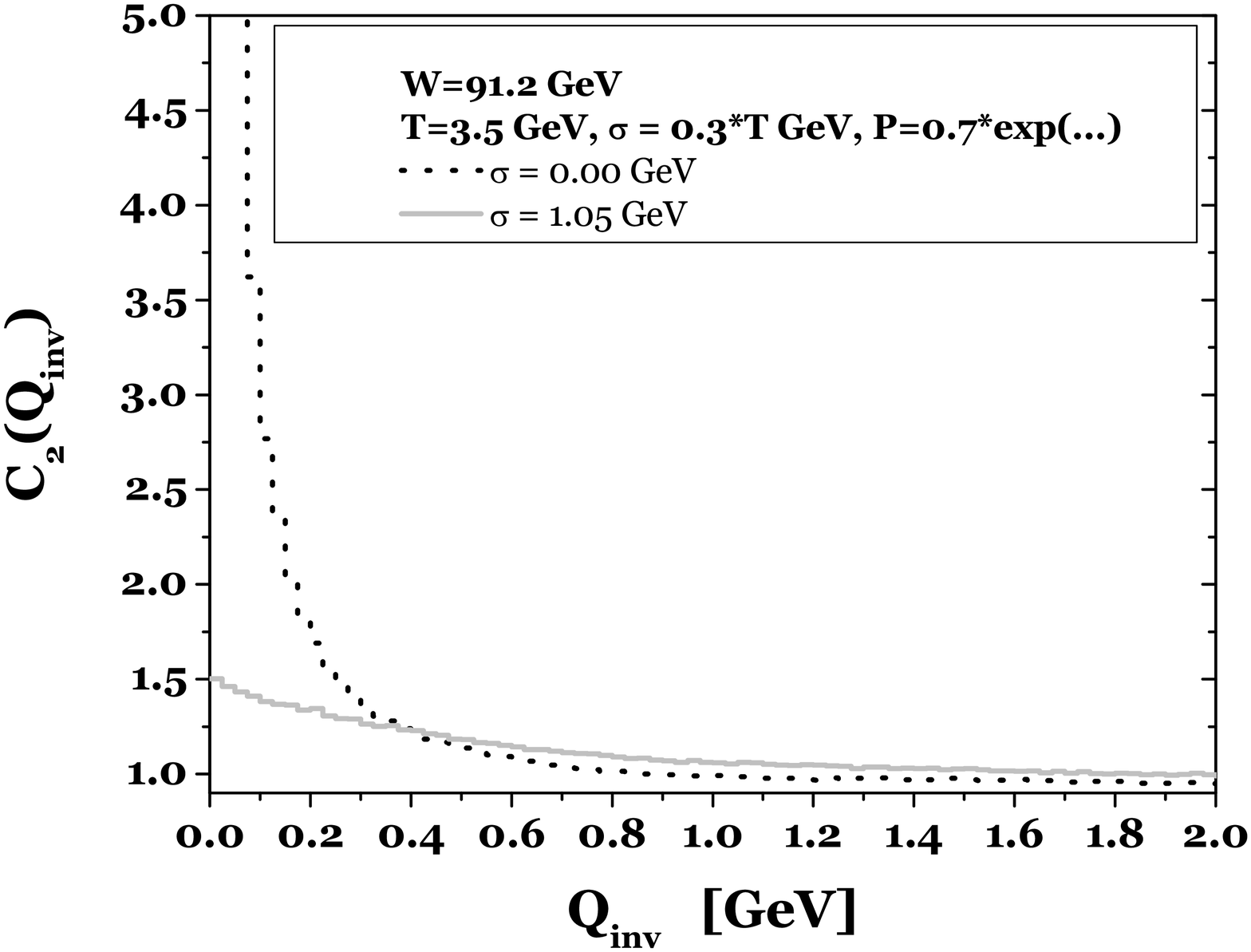}
\end{center}
\end{minipage} &
\begin{minipage}{7cm}
\begin{center}
\includegraphics[height=4.5cm,width=7cm]{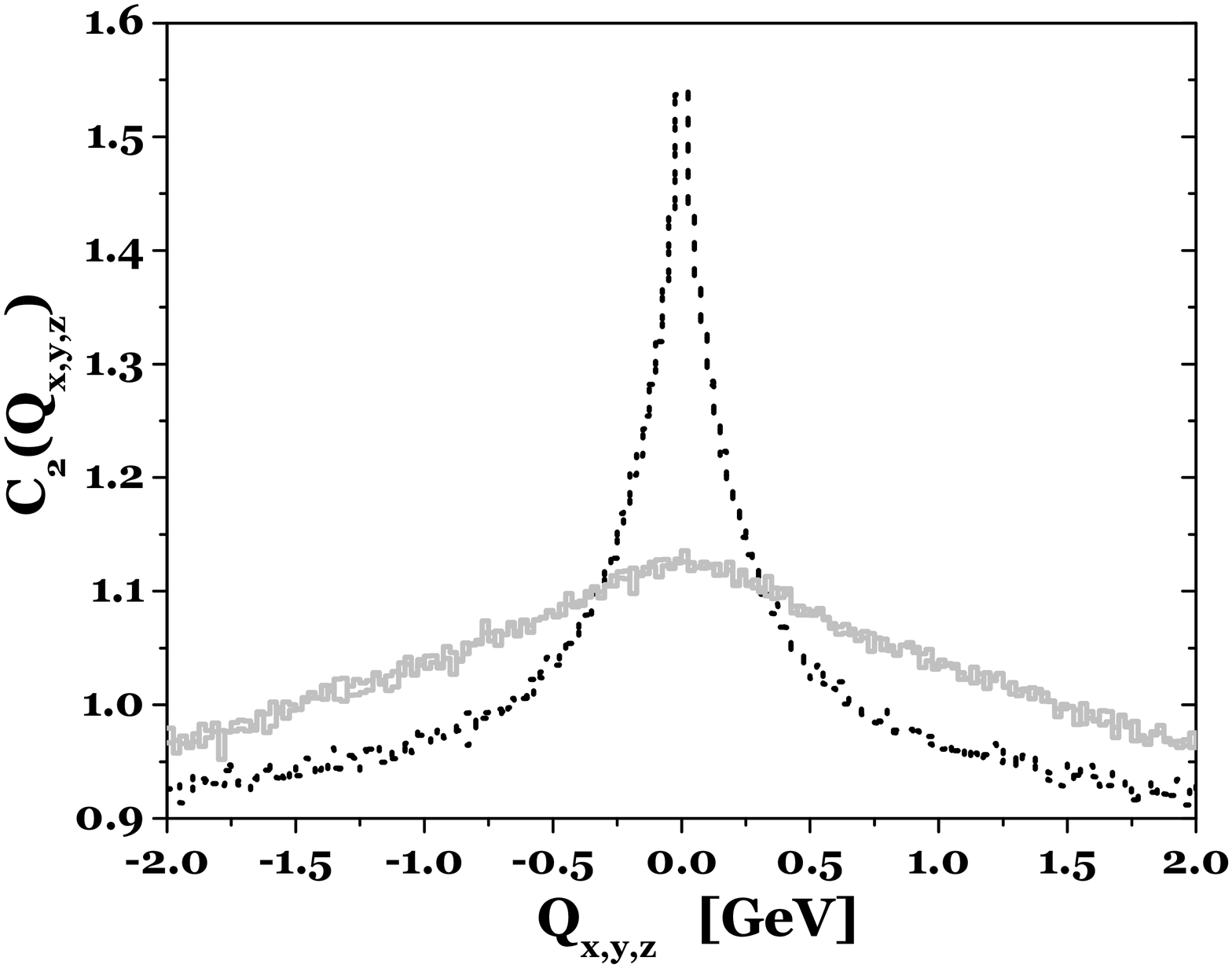}
\end{center}
\end{minipage} \\
\hline
\end{tabular} \\
\vspace{0.5cm}{\scriptsize Fig.2 Demonstration of great role played by
parameter $\sigma$ defining the energy spread of particles in the EEC.}
\end{center}

To summarize - we propose new method of numerical modelling of
hadronization events in such way as to respect the bosonic character of
produced secondaries and therefore leading to BEC. It seems to converge
in some sense to the proposition presented long time ago in \cite{ZAJC},
which was, however, in practice impossible to be implemented. The only
hope that it could work now is that in our case symmetrization is within
given EEC and not for the whole bunch of particles produced. Therefore
the number of terms involved is rather limited, whereas in \cite{ZAJC}
the whole source had to be symmetrized at once. But the effect of
including at least terms when symmetrization between, say, particles $2$
and $3$ are added to the already present symmetrization between $1$ and
$2$ and $1$ and $3$, must be carefully investigated and estimated before
any conclusion is to be reached.\\

OU is grateful for support and for the warm hospitality extended to him
by organizers of the QM2005. Partial support of the Polish State
Committee for Scientific Research (KBN) 
(grant 621/E-78/SPUB/CERN/P-03/DZ4/99 (GW)) is acknowledged.\\


\begin{thebibliography}{99}

\bibitem{HIP} Utyuzh O, Wilk G, W\l odarczyk Z (2005) Quantum Clan Model
              description of Bose-Einstein correlations. Acta Phys.
              Hung. A (HIP), in press (hep-ph/0503046).

\bibitem{NUKL} Utyuzh O, Wilk G, W\l odarczyk Z (2004) How to model BEC
               numerically? Nukleonika 49 (Supplement 2): S33-S35.

\bibitem{Purcell} Purcell E E  (1956) Nature 178 : 1447-1448.

\bibitem{Kozlov} Kozlov G A, Utyuzh O, Wilk G (2003) The Bose-Einstein
                 correlation function $C_2(Q)$ from a Quantum Field Theory
                 point of view. Phys. Rev. C68 : 024901-5.

\bibitem{ZAJC} Zajc (1987) W  Monte Carlo calculational methods for the
               generation of the evants with Bose-Einstein correlations.
               Phys. Rev. D53 : 3396-3408.


\end{thebibliography}
\end{document}